\documentclass[pre,amsmath,amssymb,twocolumn,superscriptaddress]{revtex4-2}
\usepackage{amsmath, amstext, amssymb, amsfonts, amsxtra}
\usepackage{grffile}
\usepackage{graphicx}

\newcommand{\xiamen}{Department of Physics and Key Laboratory of Low Dimensional
Condensed Matter Physics (Department of Education of Fujian Province), Xiamen
University, Xiamen 361005, Fujian, China}
\newcommand{\lanzhou}{Lanzhou Center for Theoretical Physics, Lanzhou University,
Lanzhou 730000, Gansu, China}

\begin{document}

\title{Pseudoclassical dynamics of the kicked top}

\author{Zhixing Zou}
\affiliation{\xiamen}
\author{Jiao Wang}
\affiliation{\xiamen}
\affiliation{\lanzhou}

\begin{abstract}
The kicked rotor and the kicked top are two paradigms of quantum chaos. The
notions of quantum resonance and the pseudoclassical limit, developed in the study
of the kicked rotor, have revealed an intriguing and unconventional aspect of
classical--quantum correspondence. Here, we show that, by extending these notions
to the kicked top, its rich dynamical behavior can be appreciated more thoroughly;
of special interest is the entanglement entropy. In particular, the periodic synchronization
between systems subject to different kicking strength can be conveniently understood
and elaborated from the pseudoclassical perspective. The applicability of the suggested
general pseudoclassical theory to the kicked rotor is also discussed.
\end{abstract}

\maketitle

\section{Introduction}

The study of quantum chaos, or quantum chaology~\cite{Berry89}, focuses on whether,
how, and to what extent classical chaos may manifest itself in the quantum realm.
In essence, it boils down to the general classical--quantum correspondence issue,
as insightfully pointed out by Einstein at the very early development stage of
quantum theory~\cite{Einstein17}. The quantum kicked rotor, presumably the
best known paradigm of quantum chaos, was first introduced by Casati {et al}.
in their seminal study that opened this field~\cite{Casati79}. After four decades
of investigation, the richness of this paradigmatic model appears to be surprising.
Far beyond quantum chaos, it has also been realized that this model may play
a unique role in other fundamental problems, such as Anderson localization
(transition)~\cite{AT82, AT89, AT08, AT14} and the quantum Hall effect~\cite{QH14,
QH16, QH20}. Centering around the kicked rotor, an expanded overlapping field
encompassing all these relevant problems is emerging~\cite{Garreau16}.

In contrast to its richness, another advantage of the kicked rotor lies in its
simplicity, featuring only a single point particle on a circle subject to the
stroboscopic external interaction, which makes the study of this model much
more convenient than most others. An exception is the kicked top
model~\cite{Haake1987}, which has a finite Hilbert space, so that it is even more
favorable for research. Interestingly and importantly, these two models usually
demonstrate different aspects of quantum chaos in a complementary way. With all
these advantages, they are often the first ideal candidates for probing new notions.
In recent years, interesting notions having been intensively investigated range
widely, from the out-of-time-order correlations~\cite{Maldacena2016, Rozenbaum2017,
Lerose2020}, to the dynamical entanglement~\cite{Zurek1994, Zarum1998, Wang2004,
Ruebeck2017, Bhosale2018, Lerose2020}, the non-Hermitian properties~\cite{Longhi2017,
Mudute2020}, and so on.

The dynamical entanglement is devised to capture the decoherence process
of a quantum system when being coupled to the environment. It has distinct
characteristics if the system's classical counterpart is chaotic. The kicked
top has the spin algebra symmetry and, as such, it can be regarded as a composite
of identical qubits. An additional advantage due to such a multiqubit interpretation
is that, for studying the dynamical entanglement, there is no need to introduce
the environment. It has been shown both theoretically and experimentally that,
indeed, the dynamical entanglement may serve as a diagnosis of quantum chaos
in the kicked top model~\cite{Wang2004, Chaudhury2009}.

However, as far as we know, in most previous studies of the kicked top, only
a narrower range of comparatively weak kicking strength has been investigated,
leaving its properties unexplored yet in a wider range of stronger kicking strength.
The reason might be that, for the classical kicked top, the transition from
regular to globally chaotic motion occurs at a rather weak kicking strength.
When the system is already globally chaotic, further increasing the kicking
strength would not result in any qualitatively new properties. Accordingly,
due to quantum--classical correspondence, it is reasonable to conjecture that
this would also be the case in the quantum kicked top in the semi-classical limit.
Nevertheless, as illustrated in a recent study where measures of quantum correlations
were studied~\cite{Bhosale2018}, the quantum kicked tops at certain different
kicking strengths may synchronize, in clear contrast to their classical counterparts.

In fact, it is worth noting that in the kicked rotor, the similar
classical--quantum non-correspondence phenomenon, termed quantum
resonance~\cite{Izrailev1980}, has been recognized and studied ever since
the beginning of the quantum chaos field. Later, it has also been realized
that the properties of the system when being detuned from the quantum resonance
condition can even be understood in a classical way through the so-called
pseudoclassical limit~\cite{Fishman2002A, Fishman2002B}, rather than the conventional
semi-classical limit. This reminds us to consider whether the synchronization
observed in the quantum kicked top may have any underlying connections to quantum
resonance and the pseudoclassical limit. In this work, we will provide a positive
answer to this question. In particular, we will suggest a more general scheme of
the pseudoclassical limit that involves more information of the quantum dynamics,
which allows us to successfully apply it not only to the kicked top, but also to
the kicked rotor. When being applied to the kicked rotor, the previously developed
pseudoclassical scheme is found to be a special case of the suggested one.
	
In the following, we will briefly describe the kicked top model in
Section~\ref{Sec: model} first. Next, in Section~\ref{Sec: theory}, we will discuss
the quantum resonance condition for the quantum kicked top and develop the
pseudoclassical theory. The properties of the system adjacent to the quantum
resonance condition will be discussed in detail with two illustrating cases
in Section~\ref{Sec: simulations}. In particular, the numerical studies and the
comparison with the pseudoclassical theory will be presented.
In Section~\ref{Sec: entanglement}, the properties of dynamical entanglement will
be investigated from the perspective of the pseudoclassical limit. Finally,
we will summarize our work and discuss its extension to the kicked rotor in
Section~\ref{Sec: summary}.

\section{The Kicked Top Model}
\label{Sec: model}

The Hamiltonian of the kicked top model is~\cite{Haake1987}
\begin{equation}
H=\alpha J_x+\frac{\beta}{2j}J_z^2\sum_{n=-\infty}^{\infty}\delta(t-n),
\nonumber
\end{equation}
where $J_x$, $J_y$, and $J_z$
are the angular momentum operators respecting the commutations
$\left[ J_\lambda,J_\mu \right] = i \epsilon_{\lambda \mu \nu} J_\nu$
(the Planck constant $\hbar$ is set to be unity throughout) and $j$
is an integer or half-integer related to the dimension of the Hilbert space
$N$ as $N=2j+1$. The first term in $H$ describes the procession around the
$x$ axis with angular frequency $\alpha$, while the second term accounts for
a periodic sequence of kicks with an overall kicking strength $\beta$ (the
period of kicking is set to be the unit of time). In the following, we will
restrict ourselves to the case of integer $j$, but the discussions can be
extended straightforwardly to that of half-integer $j$. Since the Hamiltonian
is time-periodic, the evolution of the system for a unit time, or one step
of evolution, can be fulfilled by applying the Floquet operator
\begin{equation}
\label{eq:Floquet}
	U=\exp\left(-i\frac{\beta}{2j}J_z^2\right)\exp\left(-i\alpha J_x\right)
\end{equation}
to the present state. Obviously, $U$ does not change under the transformation
$\beta\rightarrow\beta+4j\pi$, implying that the properties of the quantum kicked
top have a periodic dependence on the kicking strength $\beta$ of period $4j\pi$.
Thus, a better understanding of the quantum kicked top calls for investigations
covering such a period.

In the semi-classical limit $j\rightarrow \infty$, following the Heisenberg
equations,  the one-step evolution of the system reduces to the following
map~\cite{Haake1988}:
\begin{equation}
\label{eq:mapXYZ}
	\begin{split}
		X'=& X\cos[\beta (Y\sin\alpha+Z\cos\alpha)]-(Y\cos\alpha\\
		& -Z\sin\alpha)\cdot \sin[\beta(Y\sin\alpha+Z\cos\alpha)],\\
		Y'=& X\sin[\beta (Y\sin\alpha+Z\cos\alpha)]+(Y\cos\alpha\\
		& -Z\sin\alpha)\cdot \cos[\beta(Y\sin\alpha+Z\cos\alpha)],\\
		Z'=& Y\sin\alpha+Z\cos\alpha,
	\end{split}
\end{equation}
with the normalized variables $ X=J_x/j$, $Y=J_y/j$, and $Z=J_z/j$. This map
defines the classical kicked top. Physically, this map describes the process
of rotating the top along the $x$ axis for an angle of $\alpha$ first to
reach the intermediate state $(\tilde X, \tilde Y, \tilde Z)$, followed by
further rotating it around $\tilde Z$ by $\beta \tilde Z$, which is the same
as the quantum Floquet operator.

Note that the state $(X, Y, Z)$ can be viewed as a point on the surface of
a unit sphere. Therefore, it can be represented equivalently by two angles,
denoted as $\Theta$ and $\Phi$, via the coordinate transformation
$(X,Y,Z)=(\sin\Theta \cos\Phi,\sin\Theta \sin\Phi,\cos\Theta)$. For the sake
of convenience, we denote map~\eqref{eq:mapXYZ} in terms of $\Theta$ and
$\Phi$ as
\begin{equation}
\label{eq:mapTP}
	(\Theta', \Phi') = \mathcal{F} (\Theta, \Phi; \alpha, \beta),
\end{equation}
where ($\Theta'$, $\Phi'$) is the state equivalent to $(X', Y', Z')$.

In order to make a close comparison between the quantum and the classical
dynamics, we invoke the spin coherent state in the former, which has the
minimum uncertainty in a spin system. A spin coherent state centered at
$(\Theta,\Phi)$, denoted as $|\Theta,\Phi\rangle$, can be generated from
the angular momentum eigenstate $|j,j\rangle$ as
\begin{equation}
|\Theta,\Phi\rangle=
\exp\left(i\Theta\lbrack J_x \sin\Phi-J_y\cos\Phi\rbrack\right)|j,j\rangle.
\nonumber
\end{equation}

Here, $|j,j\rangle$ satisfies that $(J_x^2+J_y^2+J_z^2)|j,j\rangle=j(j+1)|j,j\rangle$
and $J_z|j,j\rangle=j|j,j\rangle$. The classical counterpart of $|\Theta,\Phi\rangle$
is the point $(\Theta,\Phi)$ on the unit sphere.

\section{The Pseudoclassical Theory}
\label{Sec: theory}

\subsection{Quantum Resonance in the Kicked Top}
The concept of quantum resonance was first introduced in the kicked rotor
model. The Floquet operator for the kicked rotor is
\begin{equation}
\label{eq:UR}
U_R=\exp\left(-i\frac{p^2}{2}T\right)\exp(-iK\cos\theta),
\nonumber
\end{equation}
where $T$ and $K$ are two parameters, $\theta$ is the angular displacement of the
rotor, and $p$ is the corresponding conjugate angular momentum. If $T=4\pi r/s$
with $r$ and $s$ as two coprime integers, except for the cases of an odd $r$ and $s=2$,
the asymptotic growth in energy is quadratic in time, corresponding to a linear
spreading of the wavepacket in the angular momentum space. This phenomenon is
referred to as ``quantum resonance''~\cite{Izrailev1980}, since it is caused by
the pure quantum effect, with no connections to the classical dynamics. Otherwise,
the energy would undergo a linear growth stage, corresponding to the diffusive
spread of the wavepacket in the angular momentum space, before it saturates
due to quantum interference, which is known as the so-called dynamical
localization~\cite{Casati95}. For the special case of an odd $r$ and $s=2$,
it follows that $U_R^2=1$. Namely, the quantum dynamics is periodic of period
two, which is referred to as ``quantum antiresonance''.

There is an implicit connection between the kicked rotor and the kicked top.
By assuming the scaling
\begin{equation}
	T=\beta/j \text{~~~and~~~} K=\alpha j,
\nonumber
\end{equation}
it has been shown that the kicked rotor emerges as a limit case of the kicked
top~\cite{Haake1988}. Given this, the notion of quantum resonance can be extended
to the kicked top by assigning the quantum resonance condition that
\begin{equation}
\beta=4j\pi \frac{r}{s}
\nonumber
\end{equation}
with coprime $r$ and $s$. Indeed, under this condition, the kicked top has similar
properties to the kicked rotor in quantum resonance. For example, for $\beta=4j\pi$,
the Floquet operator reduces to $U=\exp\left(-i\alpha J_x\right)$, implying that
the top keeps rotating at a constant rate; however, when $r$ is odd and $s=2$, we have
$U^2=1$, suggesting that the motion is periodic of period two as well.

\subsection{The Pseudoclassical Limit of the Kicked Top}

For the kicked rotor, a pseudoclassical theory has been developed to address the
quantum dynamics via a classical map, the so-called pseudoclassical limit, when
the system parameter is close to the resonance condition that $T$ is an integer
multiple of $2\pi$\linebreak {(i.e., $s=2$)}~\cite{Fishman2002A, Fishman2002B}.
In the following,  we attempt to extend this theory to the kicked top and study
its behavior for $\beta=4j\pi r/s +\delta$, where $\delta$ (incommensurate to $\pi$)
is a weak perturbation to the resonance condition that is unnecessarily limited
{to the case of $s=2$}. Suppose that the  current state is $|\Theta, \Phi \rangle$
and its classical counterpart is $\left(\Theta, \Phi \right)$; our task is to figure
out the one-step evolution result for the latter by analogy based on the quantum
evolution.

For $\beta=4j\pi r/s+\delta$, the Floquet operator~\eqref{eq:Floquet} can be
rewritten as
\begin{equation}
\label{eq:Udelta}
\begin{split}
U&=\exp\left(-i\frac{4j\pi \frac{r}{s}+\delta}{2j}J_z^2\right)\exp\left(-i\alpha J_x\right)\\
&=\exp\left(-i2\pi\frac{r}{s}J_z^2\right)\exp\left(-i\frac{\delta}{2j}J_z^2\right)\exp(-i\alpha J_x).
\end{split}
\end{equation}

Remarkably, the last two terms are exactly the Floquet operator of the kicked
top but with the kicking strength $\delta$ instead. As shown in previous studies,
when $\delta$ is small, the quantum dynamics that the last two operators represent
can be well mapped to the classical kicked top (with the kicking strength $\delta$)
in the semi-classical limit. As a consequence, the classical counterpart of the
last two operators is to map $\left(\Theta, \Phi \right)$  into the intermediate
state
\begin{equation}
\label{eq:mapdelta}
(\tilde \Theta_\delta, \tilde \Phi_\delta)
= \mathcal{F} (\Theta, \Phi; \alpha, \delta),
\end{equation}
where the subscript $\delta$ at the l.h.s. indicates the kicking strength
for the sake of clearness.

Vice versa, for the corresponding quantum evolution, due to the solid
quantum--classical correspondence in the semi--classical limit, we assume
that these two operators map the coherent state $|\Theta, \Phi \rangle$ into
that of $|\tilde \Theta_\delta, \tilde \Phi_\delta \rangle$. Then, the remaining
problem is to find out the classical counterpart of the result when the first
operator at the r.h.s. of Equation~\eqref{eq:Udelta} applies to this intermediate
state $|\tilde \Theta_\delta, \tilde \Phi_\delta \rangle$. The result is (see
the derivation in Appendix~\ref{App:A})
\begin{equation}
\label{eq:multi}
\exp \left(-i 2\pi \frac{r}{s} J_{z}^{2}\right)|\tilde \Theta_\delta, \tilde \Phi_\delta\rangle
=\sum_{l=0}^{s-1} G_l \left|\tilde \Theta_\delta, \tilde \Phi_\delta+\frac{2 \pi r}{s} l\right\rangle,
\end{equation}		
where $G_l$ is the Gaussian sum
\begin{equation}
G_l= \frac{1}{s} \sum_{k=0}^{s-1} \exp \left(-i \frac{2 \pi r}{s} k(k-l)\right).
\nonumber
\end{equation}	
	
The physical meaning of Equation~\eqref{eq:multi} is clear: the intermediate coherent
state $|\tilde \Theta_\delta, \tilde \Phi_\delta \rangle$ is mapped into $s$
coherent states located along the line of $\Theta=\tilde \Theta_\delta$, each
of which has an amplitude given by a Gaussian sum. Note that these $s$ coherent
states are not necessarily independent; some of them may correspond to the same
coherent state, if their $l$ values lead to the same angle of
$\Delta = 2\pi rl/s$ mod $2\pi$. Suppose that there are $\mathcal{N}$ different
such angles in total and denote them as $\Delta_k$, $k=1,\cdots, \mathcal{N}$;
then, Equation~\eqref{eq:multi} can be rewritten as
\begin{equation}
\label{eq:multiN}
\exp \left(-i 2\pi \frac{r}{s} J_{z}^{2}\right)|\tilde \Theta_\delta, \tilde \Phi_\delta\rangle
=\sum_{k=1}^\mathcal{N} A_k \left|\tilde \Theta_\delta, \tilde \Phi_\delta+\Delta_k\right\rangle.
\end{equation}	
		
Here, for the $k$th component coherent state, its amplitude $A_k$ is the sum
of all $G_l$ whose subscript $l$ satisfying $\Delta_k = 2\pi rl/s$ mod $2\pi$.

\begin{figure*}[!]
{\includegraphics[width=5.7cm]{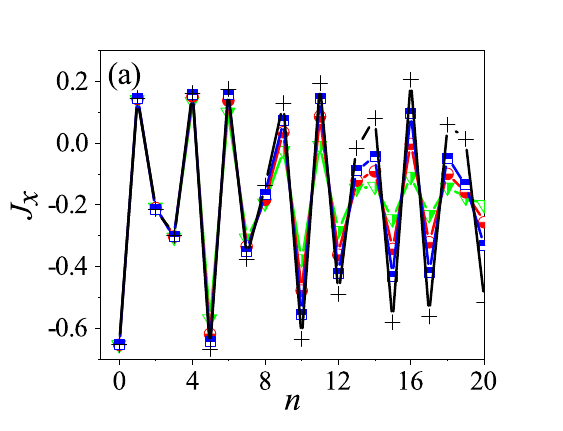}}
\hspace{-8mm}
{\includegraphics[width=5.7cm]{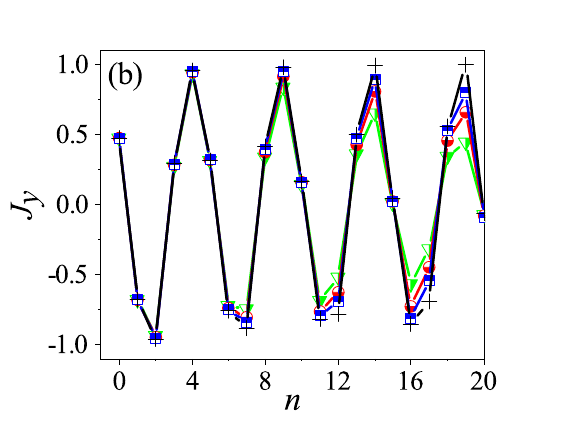}}
\hspace{-8mm}
{\includegraphics[width=5.7cm]{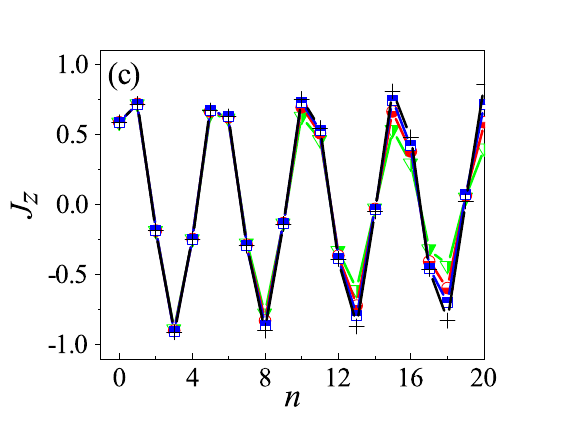}}
\caption{The time dependence of the expected value of angular momentum
$J_x$ (a), $J_y$ (b), and $J_z$ (c), respectively, for $\alpha=1$ and
$\beta=2j\pi+2$. The black crosses are for the results by the pseudoclassical
map [Eq.~\eqref{eq:caseIJ}]. The green triangles, the red circles, and the
blue squares are for the quantum results with $j=100$, $200$, and $400$,
respectively. For the initial state $(\Theta, \Phi)$ and
$|\Theta, \Phi \rangle$, $\Theta=0.8\pi$ and $\Phi=0.3\pi$.}
\label{fig1}
\end{figure*}

In the semi-classical limit $j\rightarrow\infty$, a coherent state reduces
to a point in the phase space. Given this, we can give Equation~\eqref{eq:multiN}
a classical interpretation as the following: the point $(\Theta, \Phi)$ is mapped
into a set of $\mathcal{N}$ points and meanwhile each point is associated with
a complex ``amplitude''. These two features make the situation here distinct
from the previous pseudoclassical theory for the kicked rotor, where a point
is mapped only to another point and no complex amplitude is involved. Thus,
formally, the pseudoclassical map that we seek can be expressed as
\begin{equation}
\label{eq:keyresult}
\mathcal{M}: (\Theta, \Phi) \rightarrow \{[(\tilde \Theta_\delta,
\tilde \Phi_\delta+\Delta_k); A_k], k=1,\cdots,\mathcal{N}\}.
\end{equation}		
		
This is the key result of the present work. As illustrated in the next section,
it does allow us to predict the quantum dynamics in such a pseudoclassical way.
Here, we emphasize that the amplitudes $\{A_k\}$ are crucial to this end.
Specifically, $|A_k|^2$ has to be taken as the weight of the $k$th point to
evaluate the expected value of a given observable. Moreover, the phases encoded
in these amplitudes have to be considered simultaneously to correctly trace
the quantum evolution.

\section{Applications of the Pseudoclassical Theory}
\label{Sec: simulations}

In this section, we check the effectiveness of the pseudoclassical map by comparing
its predictions with that obtained directly with the quantum Floquet operator. In
general, if a point is mapped into $\mathcal{N}>1$ points at each step, then the
number of points that we have to deal with would increase exponentially. Therefore,
in practice, it would be prohibitively difficult to apply it for any arbitrarily given
parameters. However, fortunately, for some quantum resonance parameters, $\mathcal{N}$
could be small, and, under certain conditions, e.g., if $\alpha$ is an integer multiple
of $\pi/2$, coherent cancellation may suppress the increase in the number of points
(see the second subsection below). In such cases, the application of the
pseudoclassical theory can be greatly simplified. Here, we consider two such cases
as illustrating examples, i.e., $\beta=2j\pi+\delta$ and $\beta=j\pi+\delta$,
respectively.

\subsection{Case I: $\beta=2j\pi+\delta$}

For this case, we can show (see Appendix~\ref{App:B}) that $\mathcal{N}=1$,
i.e., the pseudoclassical dynamics evolves the point $(\Theta, \Phi)$ into
another single point as
\begin{equation}
\label{eq:caseIM}
\mathcal{M}: (\Theta, \Phi) \rightarrow (\tilde \Theta_\delta,
\tilde \Phi_\delta+\pi),
\end{equation}		
with the corresponding amplitude $A_1=1$.

\begin{figure*}[!]
{\includegraphics[width=5.7cm]{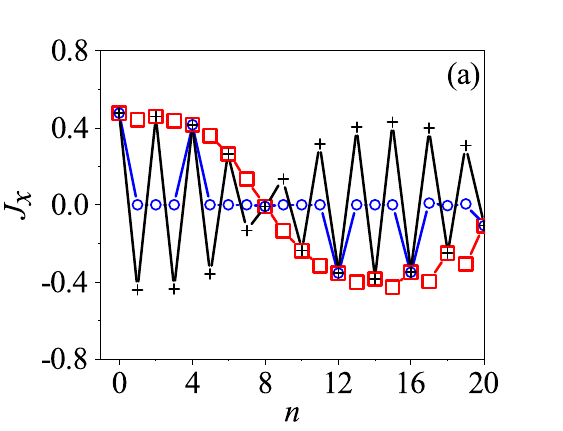}}
\hspace{-8mm}
{\includegraphics[width=5.7cm]{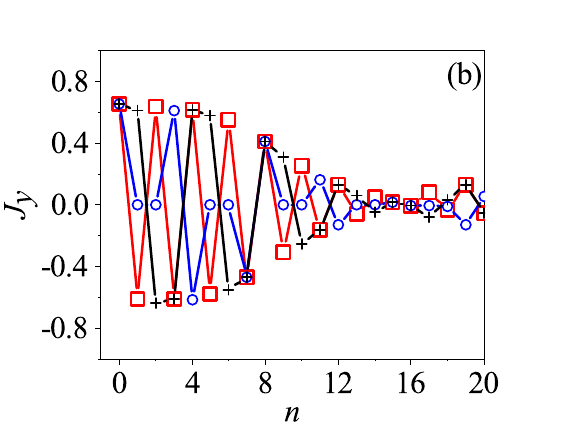}}
\hspace{-8mm}
{\includegraphics[width=5.7cm]{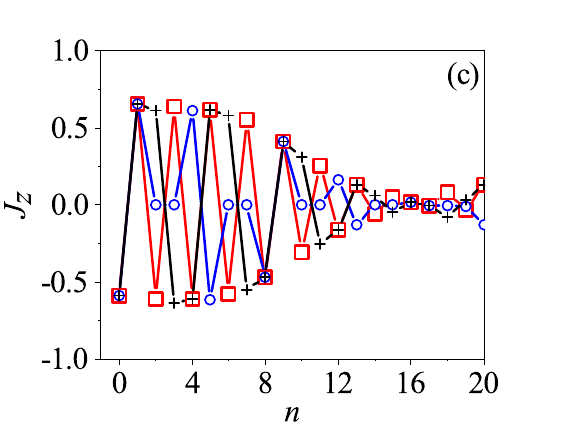}}
\caption{The same as Fig.~1 but for the quantum kicked top with
kicking strength $\beta=2j\pi+2$ (red squares), $\beta=j\pi+2$ (blue circles),
and $\beta=2$ (black crosses), respectively. Here $\alpha=\pi/2$, $j=400$, and
the initial state is $|\Theta, \Phi\rangle=|0.7\pi,0.3\pi\rangle$. }
\label{fig2}
\end{figure*}
		
With Equation~\eqref{eq:caseIM} in hand, we are ready to predict the quantum properties.
The most relevant quantities could be the expected values of angular momentums. In
Figure~\ref{fig:1}, their dependence on time is shown for a randomly chosen initial
condition. The corresponding quantum results for three different values of $j$ are
plotted together for comparison. It can be seen that the pseudoclassical results agree
very well with the quantum ones, and, as expected, as $j$ increases, the agreement
improves progressively. It shows that, indeed, the pseudoclassical limit captures
the quantum motion successfully.

The agreement illustrated in Figure~\ref{fig:1} does not depend on $\alpha$. However,
if $\alpha$ is an integer multiple of $\pi/2$, the system would have an additional
interesting property. Namely, its quantum entanglement entropy would remain synchronized
with that of the system that has a kicking strength of $\beta=\delta$
instead~\cite{Bhosale2018}. Since, for such an $\alpha$ value, the good agreement
between the pseudoclassical and the quantum evolution remains equally, we can
probe this interesting phenomenon from the pseudoclassical perspective. In fact,
by following Equation~\eqref{eq:caseIM} and taking into account the extra symmetry
introduced by such an $\alpha$ value, we can show that this synchronization
in the entanglement entropy roots in the synchronization of their dynamics
(see the following and the next section). The latter has a period of four
(two) when $\alpha$ is an odd (even) multiple of $\pi/2$. In Appendix~\ref{App:B},
the pseudoclassical dynamics is detailed for the representative example where
$\alpha=\pi/2$. For this case, in terms of the expected value of angular momentums,
denoted as $J_x$, $J_y$, and $J_z$ as well, without confusion, the connection between
these two systems at a given time $n$ can be made explicitly as follows:
\begin{align}
\label{eq:caseIJ}
J_x(n; \beta)=
		\begin{cases}
			J_x(n; \delta), & \mod(n,4)=0~\text{or}~2;\\
			- J_x(n; \delta), & \mod(n,4)=1~\text{or}~3,
		\end{cases}\nonumber\\
J_y(n; \beta)=
		\begin{cases}
			J_y(n; \delta), & \mod(n,4)=0~\text{or}~3;\\
			- J_y(n; \delta), & \mod(n,4)=1~\text{or}~2,\\
		\end{cases}\\
J_z(n; \beta)=
		\begin{cases}
			J_z(n; \delta), & \mod(n,4)=0~\text{or}~1;\\
			- J_z(n; \delta), & \mod(n,4)=2~\text{or}~3.\\
		\end{cases}\nonumber
\end{align}

Note that the angular momentum values at the l.h.s. and the r.h.s. are for the
system with kicking strength $\beta=2j\pi+\delta$ and $\delta$, respectively.

To check this prediction, we compare the numerical results of the quantum
evolution of the two systems. The results are presented in Figure \ref{fig2}, where
not only the four-step synchronization but also the details of the intermediate
states can be recognized immediately. We can also make a close comparison of
these two systems by visualizing their quantum evolution in the phase space with
the Husimi distribution~\cite{Agarwal1981}. At a given point $(\Theta,\Phi)$
in the phase space, the Husimi distribution $P(\Theta,\Phi)$ is defined as
the expectation value of the density matrix $\rho$ with respect to the
corresponding spin coherent state, i.e.,
\begin{equation}\label{eq:13}
P(\Theta,\Phi)=\frac{2j+1}{4\pi}\langle \Theta,\Phi|\rho|\Theta,\Phi\rangle.
\nonumber
\end{equation}

\begin{figure}[!b]
{\includegraphics[width=4.3cm]{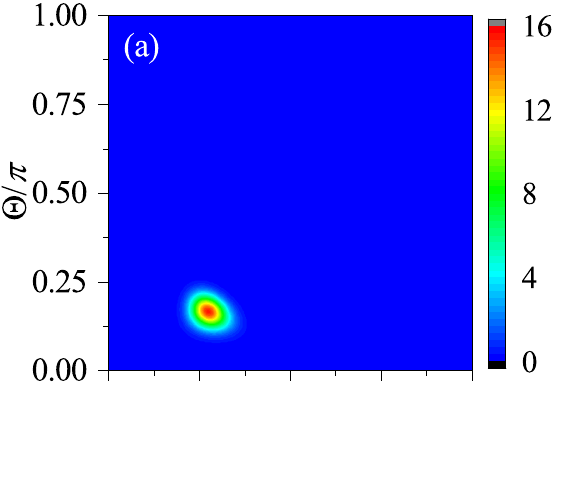}}
\hspace{-8mm}
\vspace{-6mm}
{\includegraphics[width=4.3cm]{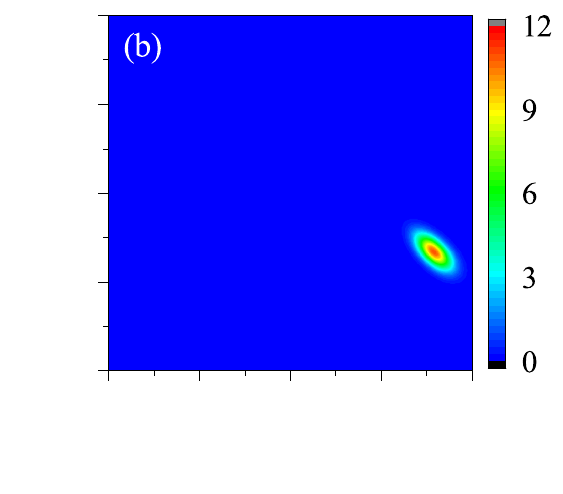}}\\
{\includegraphics[width=4.3cm]{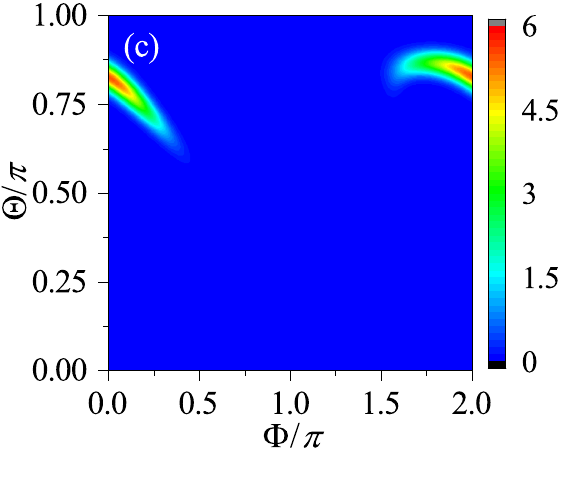}}
\hspace{-8mm}
{\includegraphics[width=4.3cm]{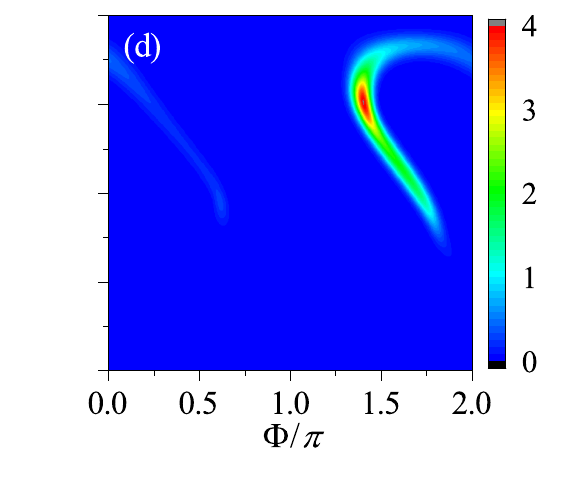}}
\caption{The Husimi distribution for $\beta=2$ at the time $n=1$ (a), $n=2$ (b),
$n=4$ (c) and $n=8$ (d), respectively. Here $\alpha=\pi/2$, $j=100$, and the
initial state is $|\Theta,\Phi\rangle=|\pi/2,\pi/3\rangle$.}
\label{fig3}
\end{figure}

\begin{figure}[!b]
{\includegraphics[width=4.3cm]{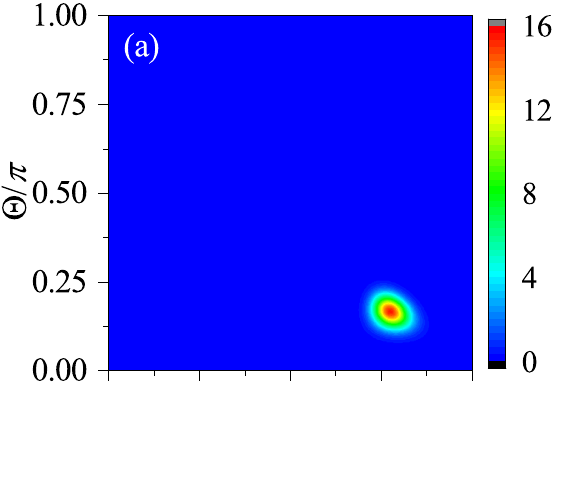}}
\hspace{-8mm}
\vspace{-6mm}
{\includegraphics[width=4.3cm]{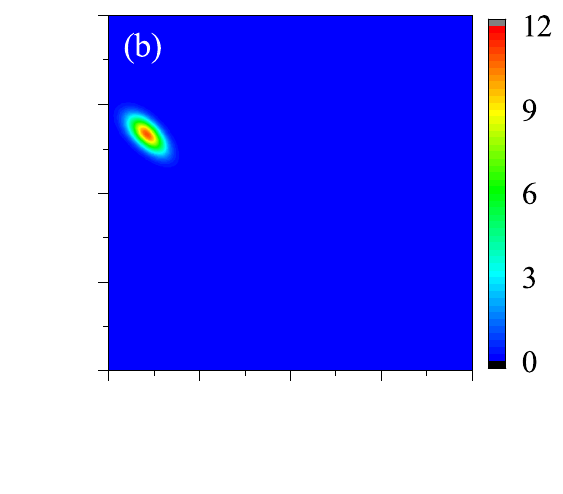}}\\
{\includegraphics[width=4.3cm]{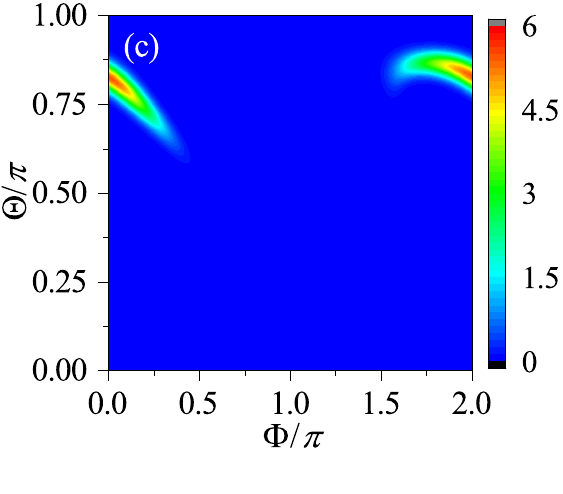}}
\hspace{-8mm}
{\includegraphics[width=4.3cm]{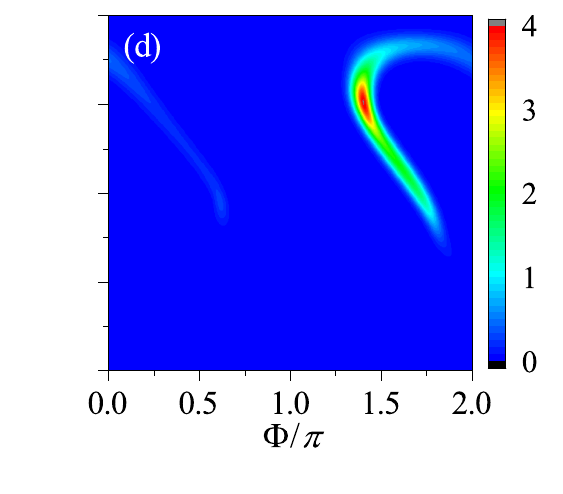}}
\caption{
The same as Fig.~3 but for $\beta=2j\pi+2$ instead (other parameters remain
unchanged).}
\label{fig4}
\end{figure}

The results for $\beta=2$ and $\beta=2j\pi+2$ at four different times are
shown in Figures~\ref{fig3} and \ref{fig4}, respectively. It can be seen that, when $n=1$,
the centers of the two wavepackets only differ by an angle of $\pi$ in $\Phi$,
while, when $n$=2, they become symmetric with respect to
$(\Theta, \Phi)=(\pi/2, \pi)$. When $n$=4 and $8$, the two wavepackets are
indistinguishable, which is a sign that the two systems are synchronized.

Obviously, all these numerical checks have well corroborated the effectiveness
of our pseudoclassical analysis.

\begin{figure}[!t]
{\includegraphics[width=4.3cm]{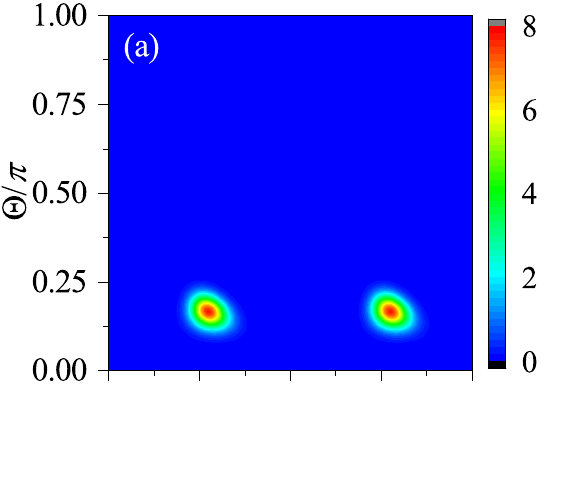}}
\hspace{-8mm}
\vspace{-6mm}
{\includegraphics[width=4.3cm]{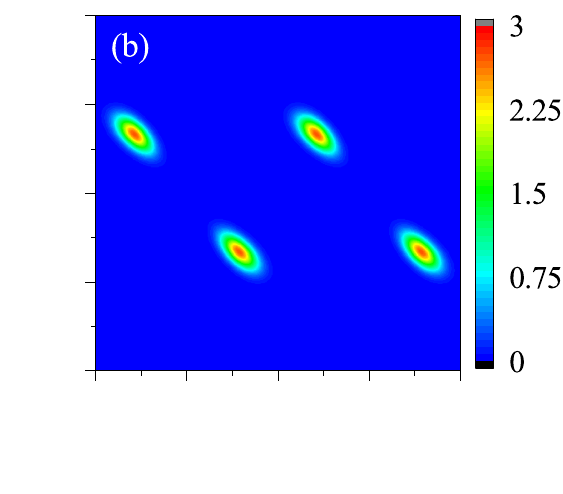}}\\
{\includegraphics[width=4.3cm]{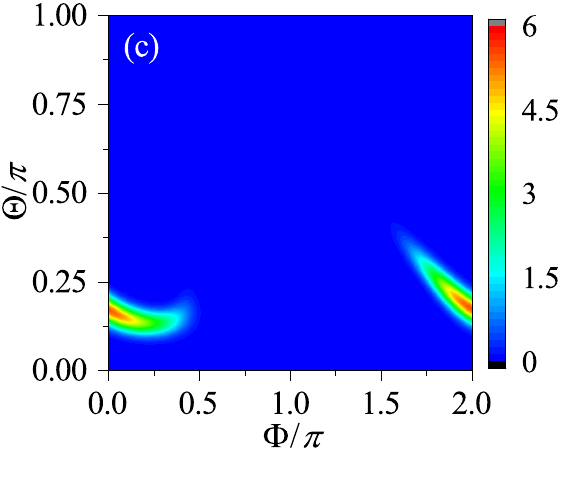}}
\hspace{-8mm}
{\includegraphics[width=4.3cm]{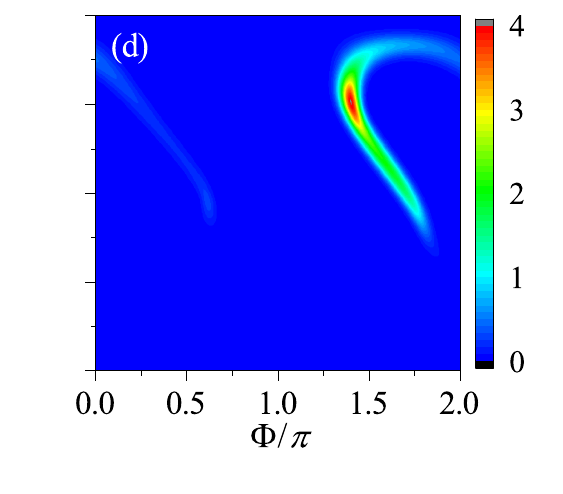}}
\caption{
The same as Fig.~3 and Fig.~4 but for $\beta=j\pi+2$. All other parameters are
the same as in the former two.}
\label{fig5}
\end{figure}

\subsection{Case II: $\beta=j\pi+\delta$}

Now, let us consider a more complex case, i.e., $\beta=j\pi+\delta$. For this
case, $\mathcal{N}=2$ and the pseudoclassical map is (see Appendix \ref{App:C})
\begin{align}
\label{eq:caseIIM}
\mathcal{M}: (\Theta, \Phi) \rightarrow
		\begin{cases}
			(\tilde \Theta_\delta, \tilde \Phi_\delta+\pi);~~A_1,\\
			(\tilde \Theta_\delta, \tilde \Phi_\delta);~~A_2,
		\end{cases}
\end{align}
with $A_1=(1+i)/2$ and $A_2=(1-i)/2$. It implies that, after each step, a point
will be mapped into two points at the same probability but with different phases.
This map looks simple, but as $\mathcal{N}=2$, if we use it to predict the quantum
evolution, the points will proliferate in time so that, in practice, we can trace
the quantum motion for a few steps only. Interestingly, this fact might explain
why the quantum motion would be complicated from a new perspective.

Nevertheless, for some special values of $\alpha$, due to the coherent effect, the
newly generated points after a step of iteration may overlap and cancel each other out,
making the number of points increase more slowly. An intriguing example is that
discussed in the previous subsection, i.e., where $\alpha$ is an integer multiple of
$\pi/2$. Again, for such an $\alpha$ value, the system is brought to synchronization
with the system of kicking strength $\beta=\delta$ as well, but with instead a
period of eight (four) if $\alpha$ is an odd (even) multiple of $\pi/2$. To be
explicit, for $\alpha=\pi/2$, the connections between the two systems are presented
in Appendix \ref{App:C}. In terms of the expected value of angular momentums, we
have
\begin{align}
\label{eq:caseIIJ}
J_x (n; \beta) & =
			\begin{cases}
				J_x(n; \delta), &\quad \mod(n,8)=0~\text{or}~4;\\
				0, & \quad\quad\text{else},
			\end{cases}\nonumber\\
J_y (n; \beta) & =
			\begin{cases}
				J_y(n; \delta), & \mod(n,8)=0~\text{or}~7;\\
				- J_y(n; \delta), & \mod(n,8)=3~\text{or}~4;\\
				0, & \quad\text{else},
			\end{cases}\\
J_z (n; \beta) & =
			\begin{cases}
				J_z(n; \delta), & \mod(n,8)=0~\text{or}~1;\\
				- J_z(n; \delta), & \mod(n,8)=4~\text{or}~5;\\
				0, & \quad\text{else}.
			\end{cases}\nonumber
\end{align}

The simulation results of the quantum angular momentums for  $\beta=j\pi+\delta$
are shown in Figure~\ref{fig2} as well; they support this derivation convincingly.

Based on the pseudoclassical dynamics, we find that the initial point will be
mapped into two and then four points after the first and the second iteration,
respectively. However, after the third iteration, the points do not become
eight as expected; rather, these eight points can be divided into four pairs
and the two points in each pair overlap with each other. Moreover, two of these
four points disappear in effect as the resultant total amplitude for each of
them turns out to be zero (see Appendix~\ref{App:C}). Thus, only two points remain,
and after the fourth iteration, these two points further merge into one. As a
consequence, the number of points varies in time with a period of four.  The
results for the Husimi distribution given in Figure~\ref{fig5} are in good agreement with
this analysis. Comparing this with the results for $\beta=\delta$ in Figure~\ref{fig3}, we can
see that after the fourth iteration, there is only one wavepacket of the same
shape in both cases, but their positions are different. Only after the eighth
iteration, the two wavepackets are identical, which explains why the
synchronization period should be eight.

\begin{figure}[!t]
\includegraphics[width=8.6cm]{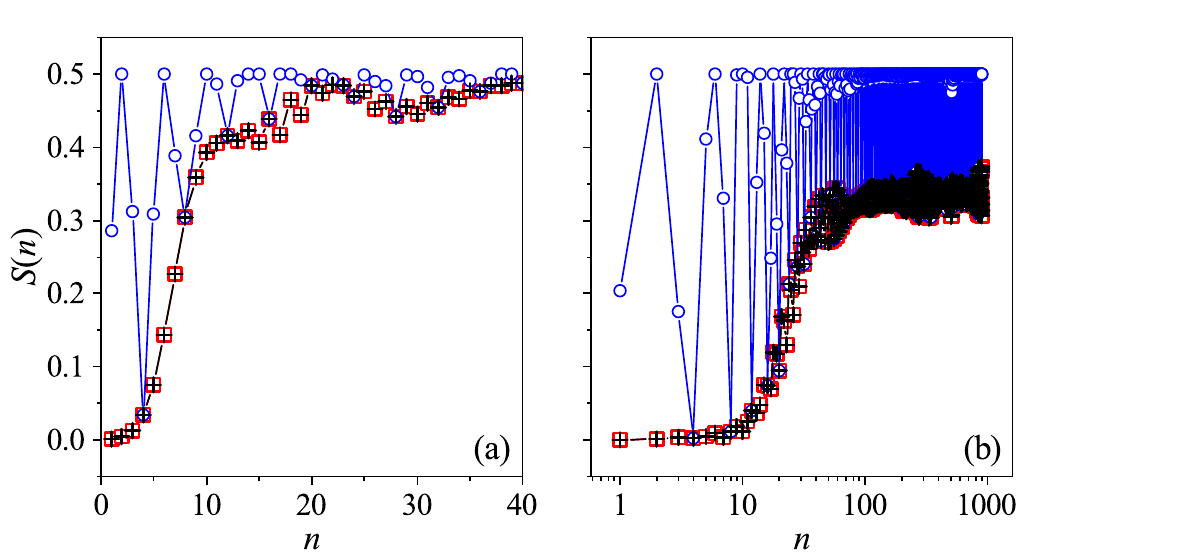}
\caption{The linear entropy as a function of time for the initial condition
$|\Theta, \Phi\rangle=|0.7\pi,0.3\pi\rangle$ (a) and $|0.7\pi,0.6\pi\rangle$
(b), respectively. The classical counterparts of these two states are in the
chaotic and regular region of the phase space, respectively, for the classical
kicked top of $\beta=2$. In both panels, the red squares, the blue circles,
and the black crosses are for, respectively, $\beta=2j\pi+2$, $\beta=j\pi+2$,
and $\beta=2$. For all the cases $\alpha=\pi/2$ and $j=400$.}
\label{fig6}
\end{figure}	

\section{The Dynamical Entanglement}
\label{Sec: entanglement}

\begin{figure*}[!t]
{\includegraphics[width=6cm]{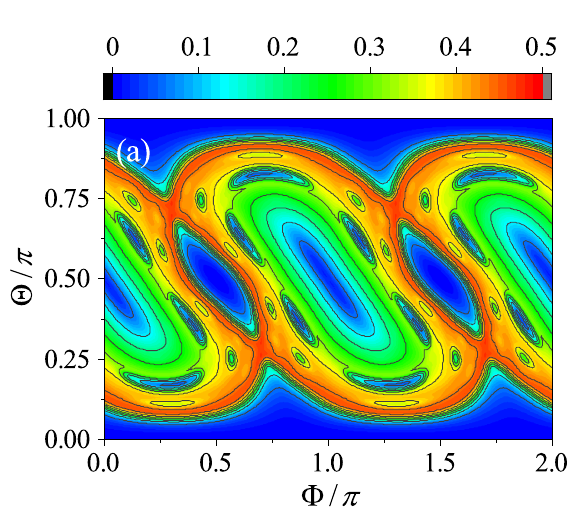}}
\hspace{-8mm}
{\includegraphics[width=6cm]{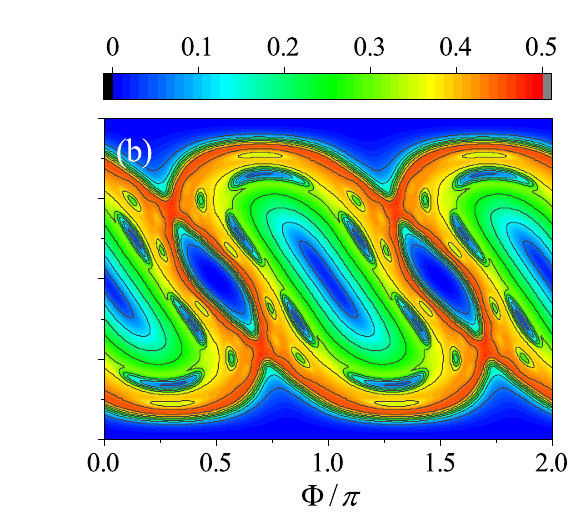}}
\hspace{-8mm}
{\includegraphics[width=6cm]{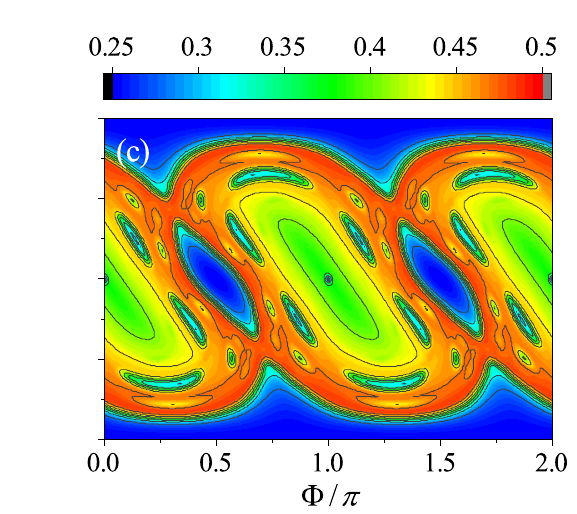}}
\caption{The contour plot of the time averaged entanglement entropy, $S_\tau$
for kicking strength  $\beta=2$ (a), $\beta=2j\pi+2$ (b), and $\beta=j\pi+2$
(c), respectively. Here $\alpha=\pi/2$, $j=400$, $\tau=300$, and a grid of
$201\times 201$ initial coherent states are simulated for each case.}
\label{fig7}
\end{figure*}

The dynamical entanglement of the quantum kicked top has been studied carefully
in recent years. We discuss this issue in this section by taking advantage of
the pseudoclassical results obtained in the previous section.

{For the quantum kicked top, its momentum can be represented by $2j$ qubits,
or a collection of $2j$ spin-1/2 identical particles. If the initial state of the
system is symmetric under permutations for identical qubits, this permutation
symmetry will be preserved, as it is respected by the action of the Floquet
operator of the kicked top. As a consequence, the expected spin value for any
single qubit of these $2j$ identical qubits is
$ s_{\gamma} = J_\gamma /(2j)$, where $J_\gamma$ is the expected momentum value
of the top and $\gamma=x, y,$ and $z$ ~\cite{Ghose2008}. }

{On the other hand,} though various bipartite entanglement measures have
been suggested, it has been shown that they are qualitatively equivalent. Thus, we
adopt the measure considered the most frequently, i.e., the bipartite entanglement
between any qubit and the subsystem made up of the remaining $2j-1$ qubits.
{This entanglement is usually quantified by computing the linear entropy
$S=1-\mathrm{Tr}(\rho_1^2)$, where $\rho_1$ denotes the reduced density operator
for a single qubit. As $\rho_1$ is a $2\times 2$ operator, it can be expressed
as $\rho_1=1/2+\sum_\gamma s_\gamma \sigma_\gamma$, given the expected spin value
$s_\gamma$. Here, $\sigma_\gamma$ is the Pauli operator.  By substituting
$s_{\gamma} = J_\gamma /(2j)$, we have
$\rho_1 = 1/2+\sum_\gamma J_\gamma \sigma_\gamma/(2j)$
in terms of $J_\gamma$ instead~\cite{Ghose2008}. It follows that}
\begin{equation}
\label{equ:S}
S=\frac{1}{2}\left( 1-\frac{[J_x] ^2+ [J_y] ^2+[J_z] ^2}{j^2}\right),
\end{equation}
which has a well-defined classical counterpart and is easy to compute numerically.
Here, $[J_\gamma]^2$ represents the square of the expected value of angular
momentum $J_\gamma$.

For the two cases close to the quantum resonance condition discussed in the
previous section, we can see immediately how their linear entropy is related to
that of the case $\beta=\delta$ based on the pseudoclassical analysis. First,
for $\beta=2j\pi+\delta$, from Equation~\eqref{eq:caseIJ}, we have that
$[J_\gamma(n;\beta)]^2=[J_\gamma(n;\delta)]^2$ at any time $n$; hence, $S(n)$
must coincide with that for the system of $\beta=\delta$ throughout. However, for
$\beta=j\pi+\delta$, from Equation~\eqref{eq:caseIIJ}, we know that
$\sum_\gamma [J_\gamma(n;\beta)]^2=\sum_\gamma [J_\gamma(n;\delta)]^2$
only when $\mod(n,8)=0$ or $4$. Therefore, we may expect a synchronization
of period four in the linear entropy. In addition, from Equation~\eqref{eq:caseIIJ},
we also know that $\sum_\gamma [J_\gamma(n;\beta)]^2=0$
when $\mod(n,8)=2$ or $6$, suggesting that the linear entropy should reach
its maximal value repeatedly in a period of four as well. As for the case
$\beta=\delta$ itself, because the quantum motion can be approximated by
the semi-classical limit if $\delta$ is small, the time dependence of the
linear entropy can be predicted based on the classical dynamics~\cite{Lerose2020}.
In a chaotic region of the phase space, $S(n)$ should increase linearly before
saturation; otherwise, it proceeds in a logarithmic law that features the regular motion.
The numerical results of the linear entropy for these three cases and two
representative initial states are presented in Figure~\ref{fig6}, which meet all these
expectations.

\begin{figure}[!]
\includegraphics[width=8.6cm]{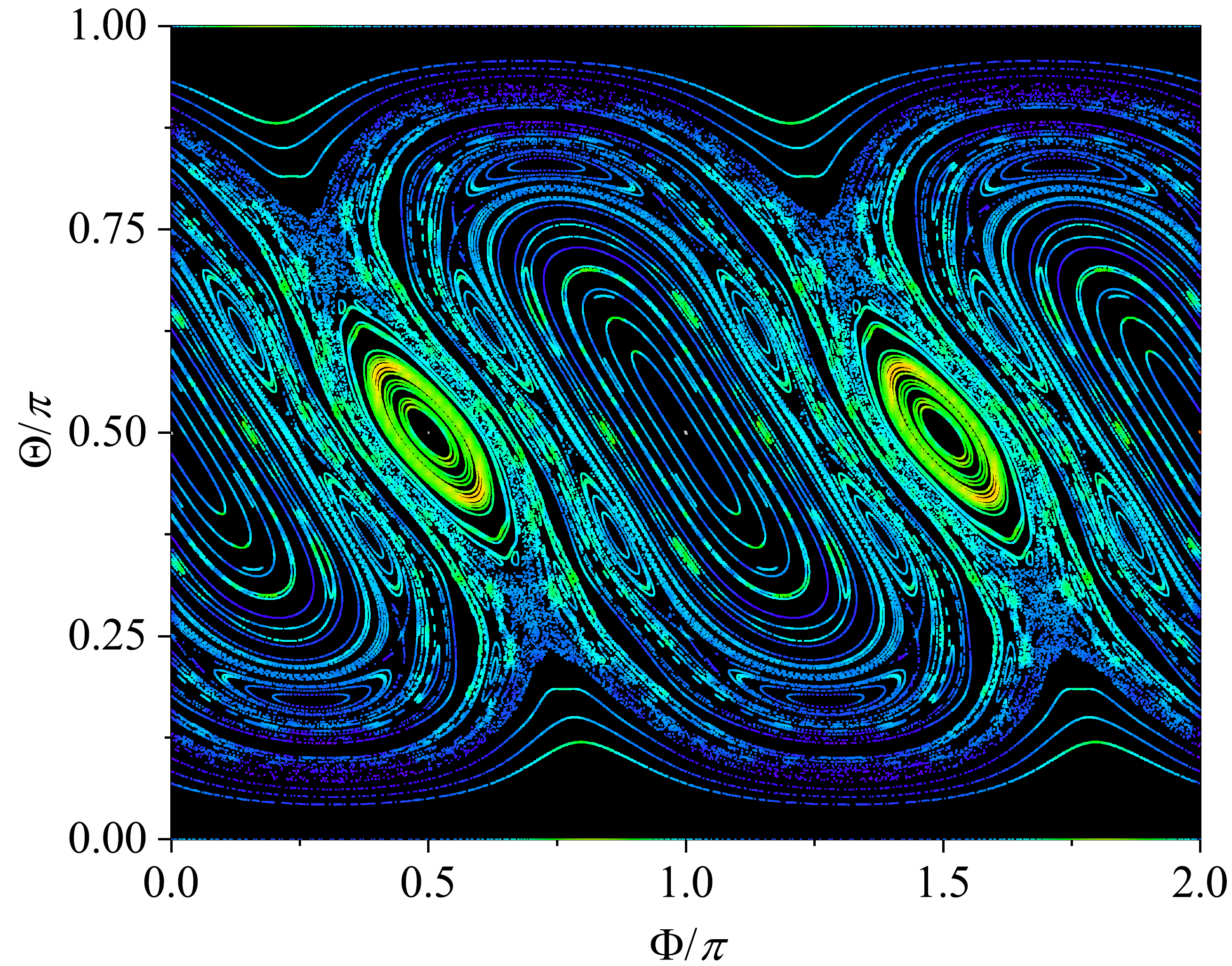}
\caption{The phase space portrait of the classical kicked top with
$\alpha=\pi/2$ and $\beta=2$.}
\label{fig8}
\end{figure}

Another interesting and related quantity that has been intensively investigated
is the time-averaged entanglement entropy. If the semi-classical limit exists,
it is used to estimate the equilibrium value that the corresponding classical system
tends to. For the linear entropy, it is defined as
\begin{equation}
\label{equ:17}
S_\tau=\frac{1}{\tau}\sum_{n=1}^{\tau}S(n),
\end{equation}
where $\tau$ is a sufficiently long time. Interestingly, it was found that the contour
plot of $S_\tau$ can well capture the characteristics of the phase space portrait of
the corresponding classical system~\cite{Zarum1998,Wang2004}. The reason is that,
given the semi-classical limit, for all initial coherent states centered on the
same classical trajectory, by definition, $S_\tau$ should be the same in the long
average time limit. As such, $S_\tau$ can be used to distinguish the trajectories.
As an illustration, in Figure~\ref{fig7}a, the contour plot of $S_\tau$ for the case $\beta=2$,
where the semi-classical limit holds well, is shown. The corresponding phase
space portrait of its classical counterpart is shown in Figure~\ref{fig8}. The similarity
between them is easy to recognize.

For the two cases close to the quantum resonance condition, the semi-classical
limit breaks. However, as the pseudoclassical limit exists, it allows us to
use $S_\tau$ to probe the corresponding pseudoclassical systems. In Figure~\ref{fig7}b,
the contour plot for $\beta=2j\pi+2$ is shown. It is identical to that for
$\beta=2$, because the two cases share the same linear entropy at every step.
However, it is worth noting that, although, from Equation~\eqref{eq:caseIM},
we know that any trajectory of the pseudoclassical system for $\beta=2j\pi+\delta$
is related to one of the semi-classical system for $\beta=\delta$, they are
not located at the same positions in their respective phase spaces (see Figures~\ref{fig3}
and \ref{fig4}, for example). This suggests that $S_\tau$ is still equally helpful for
obtaining an overall sense of the pseudoclassical dynamics, but some details
could be missed inevitably by definition.

More interesting is the case for $\beta=j\pi+\delta$. For the pseudoclassical
dynamics, the concept of the conventional trajectory does not apply any longer,
because, at a given time, the number of points in the phase space can be multiple,
and they have their respective complex amplitudes. Regardless of this fact,
as Figure~\ref{fig7}c shows, $S_\tau$ works well again for schematizing the pseudoclassical
dynamics. For example, due to the periodic synchronization with the case of
$\beta=\delta$, we may expect that the chaotic and regular regions of the former
are exactly those of the latter, respectively. This is indeed the case, which
can be seen by comparing Figure~\ref{fig7}a,c. Alternatively, if we compute $S_\tau$ by
taking the average of $S(n)$ once every four steps (the period of synchronization),
the results should be the same exactly for both cases.

\section{Summary and discussion}
\label{Sec: summary}

In summary,  we have introduced the quantum resonance condition into the kicked
top model. In order to study the behavior of the quantum kicked top {{ detuned from}}
the quantum resonance condition, we have established the corresponding
pseudoclassical theory. By analytical and numerical studies, we have shown that
this theory is effective. In particular, when being applied to discuss the
dynamical entanglement, the properties of the quantum kicked top are successfully
predicted based on the pseudoclassical dynamics. Our results also suggested that
the time-averaged entanglement entropy is still a powerful tool for grasping the
pseudoclassical dynamics.

The suggested pseudoclassical scheme is distinct from the one originally introduced
in the kicked rotor that works only near the special quantum resonance
condition that $T$ is an integer multiple of $2\pi$. To make this explicit,
let us extend our scheme to the kicked rotor and compare it with the original
pseudoclassical limit. The result is similar to
Equation~\eqref{eq:keyresult}:
\begin{equation}
\mathcal{M}: (p, \theta) \rightarrow \{[(\tilde p_\delta,
\tilde \theta_\delta+\Delta_k); A_k], k=1,\cdots,\mathcal{N}\},
\end{equation}		
but, here, $(p, \theta)$ is the classical counterpart and the center of the
coherent state $|p, \theta\rangle$ for the kicked rotor, and
$(\tilde p_\delta, \tilde \theta_\delta)$
is that $(p, \theta)$ is mapped to by the classical kicked rotor dynamics
with the kicking strength $\delta K$. For $T=2k\pi+\delta$ ($k$ is an integer),
it gives exactly the original pseudoclassical result. However, this scheme
also applies when the system is close to other quantum resonances, making
it more general than the original one.

In practice, the main challenge for applying this scheme is the same as that
encountered in the kicked top, i.e., the rapid proliferation of phase space
points. However, if we introduce an additional symmetry into the system, i.e.,
the translation invariance for $\theta \rightarrow \theta+2\pi/w$, where
$w=s/2$ for an even $s$ and $w=s$ for an odd $s$, respectively, it is found that
the coherent cancellation mechanism works efficiently so that the proliferation
can be greatly suppressed. The translation invariance can be fulfilled by
replacing the potential $\cos\theta$ in the Floquet operator $U_R$
with $\cos(w\theta)$. Such a favorable property makes the kicked rotor even
more advantageous than the kicked top for demonstrating the pseudoclassical
dynamics. A detailed discussion will be published elsewhere~\cite{Zou22}.

With the pseudoclassical theory in hand, some interesting problems could be
investigated further. For example, {it may be applied to study the entanglement
in the kicked top with weak measurements of one or several qubits~\cite{Schomerus22}
to identify the measurement effect from a different perspective.} Moreover,
the proliferation of the phase space points is in clear contrast with the
conventional classical dynamics. As it is inherited from the quantum evolution,
$\log{\mathcal{N}}$ might be taken as a complexity measure of the quantum dynamics.
For some previously studied problems in the double kicked rotor and top, such as
Hofstadter's butterfly spectrum~\cite{JB08A, JB09A, JB09B} and exponential
and superballistic wavepacket spreading~\cite{JB11, FP16}, it may help us to gain
a deeper understanding. For some frontier topics mentioned in the Introduction,
this theory may find applications as well.

\section*{Acknowledgements}

This work is supported by the National Natural Science Foundation of China
(Grants No. 12075198 and No. 12047501).


\appendix
\widetext

\section{Derivation of Eq.~(6)}
\label{App:A}

First, we expand the coherent state $|\tilde \Theta_\delta, \tilde \Phi_\delta \rangle$
over the eigenstates $\left\lbrace |m\rangle\right\rbrace$ of $J_z$, i.e.,
$|\tilde \Theta_\delta, \tilde \Phi_\delta \rangle=\sum_{m=-j}^{j} c_m |m\rangle$.
Applying the operator $\exp\left(-i\frac{2\pi r}{s}J_z^2\right)$ to both sides,
it results in
\begin{equation}\label{eq:A1}
\begin{split}
\exp \left(-i \frac{2 \pi r}{s} J_{z}^{2}\right)|\tilde \Theta_\delta,
\tilde \Phi_\delta\rangle =\sum_{k=0}^{s-1}\exp\left(-i\frac{2\pi r}{s}
k^2\right) \sum_{\mod(m,s)=k} c_m|m\rangle.
\end{split}
\end{equation}
Note that by the translation operator $\exp(-iJ_z \phi)$, coherent state
$|\tilde \Theta_\delta, \tilde \Phi_\delta \rangle$ is shifted into
$|\tilde \Theta_\delta, \tilde \Phi_\delta + \phi \rangle$, we thus have
\begin{equation}\label{eq:A2}
\begin{split}
\left|\tilde \Theta_\delta, \tilde \Phi_\delta+\frac{2 \pi r}{s} l\right\rangle
&=\exp \left(-i J_{z} \frac{2 \pi r}{s} l\right)|\tilde \Theta_\delta,
\tilde \Phi_\delta\rangle\\
&=\sum_{k=0}^{s-1} \exp \left(-i \frac{2 \pi k r}{s} l\right)
\sum_{\bmod (m, s)=k} c_{m}|m\rangle
\end{split}
\end{equation}
by setting $\phi=2\pi lr/s$, where $l$ is an integer. Next, multiplying both
sides with $\exp(i 2\pi \lambda lr/s)$, where $\lambda$ is an integer,
$0\le \lambda \le s-1$, and taking summation over $l$ from $l=0$ to $s-1$,
we can obtain that
\begin{equation}\label{eq:A3}
\sum_{l=0}^{s-1} \frac{1}{s}
\exp \left(i \frac{2 \pi r}{s} \lambda l\right)\left|\tilde \Theta_\delta,
\tilde \Phi_\delta+\frac{2 \pi r}{s} l\right\rangle
=\sum_{\bmod (m, s)=\lambda} c_{m}|m\rangle.
\end{equation}
Finally, by replacing $\lambda$ in Eq.~\eqref{eq:A3} with $k$ and substituting
this equation into Eq.~\eqref{eq:A1}, we have the result of Eq.~\eqref{eq:multi}.

\section{Pseudoclassical dynamics for $\beta=2j\pi+\delta$}
\label{App:B}

For this case, $r=1$ and $s=2$; we have $G_0=0$, $G_1=1$, $A_1=1$, and
$\mathcal{N}=1$, so that the pseudoclassical map Eq. \eqref{eq:keyresult}
reduces to
\begin{equation}
\mathcal{M}: (\Theta, \Phi) \rightarrow (\tilde \Theta_\delta, \tilde \Phi_\delta+\pi).
\end{equation}	
Namely, a point is mapped into another with unity amplitude.

On the other hand, note that for the particular case that $\alpha=\frac{\pi}{2}$,
the classical map given by Eq. \eqref{eq:mapTP} has the following properties:
\begin{equation}
\label{eq:alpha-half-pi}
    \begin{split}
        (\pi-\Theta',\pi-\Phi')=&\mathcal{F}(\Theta,\Phi+\pi,\frac{\pi}{2},\beta),\\
        (\Theta',\Phi'+\pi)=&\mathcal{F}(\pi-\Theta,\pi-\Phi,\frac{\pi}{2},\beta),\\
        (\pi-\Theta',2\pi-\Phi')=&\mathcal{F}(\pi-\Theta,2\pi-\Phi,\frac{\pi}{2},\beta).\\
    \end{split}
\end{equation}

If we use $(\Theta(0), \Phi(0))$ to denote the initial state and
$(\tilde \Theta_\delta(n),\tilde \Phi_\delta(n))$ the state after $n$ kicks
following the classical map with kicking strength $\delta$, i.e.,
\begin{equation}
(\tilde \Theta_\delta(n),\tilde \Phi_\delta(n))=\mathcal{F}^n
(\Theta_\delta(0), \Phi_\delta(0), \frac{\pi}{2},\delta),
\end{equation}
we can write down the results of the pseudoclassical map step by step as
follows:

\boldmath $n=0\rightarrow n=1$ \unboldmath:
\begin{equation*}
(\Theta(0),\Phi(0)) \rightarrow  (\tilde \Theta_\delta(1), \tilde \Phi_\delta(1)+\pi);
\end{equation*}
	
\boldmath $n=1\rightarrow n=2$ \unboldmath:
\begin{equation*}
(\tilde \Theta_\delta(1), \tilde \Phi_\delta(1)+\pi) \rightarrow
(\pi-\tilde \Theta_\delta(2), 2\pi-\tilde \Phi_\delta(2));
\end{equation*}
	
\boldmath $n=2\rightarrow n=3$ \unboldmath:
\begin{equation*}
(\pi-\tilde \Theta_\delta(2), 2\pi-\tilde \Phi_\delta(2))\rightarrow
(\pi-\tilde \Theta_\delta(3), \pi-\tilde \Phi_\delta(3));
\end{equation*}
	
\boldmath $n=3\rightarrow n=4$ \unboldmath:
\begin{equation*}
(\pi-\tilde \Theta_\delta(3), \pi-\tilde \Phi_\delta(3))\rightarrow
(\tilde \Theta_\delta(4), \tilde \Phi_\delta(4)).
\end{equation*}
At each step, the amplitude is unity. It shows that, after every four steps,
the pseudoclassical dynamics coincides with the classical dynamics of kicking
strength $\delta$. Based on these results, we can write down the results given
in Eq.~\eqref{eq:caseIJ} straightforwardly.

\section{Pseudoclassical dynamics for $\beta=j\pi+\delta$}
\label{App:C}

For this case, $r=1$ and $s=4$; we have $G_1=0$,
$G_2=\frac{\sqrt{2}} {2}\exp(i\frac{\pi}{4})$, $G_3=0$, and
$G_4=\frac{\sqrt{2}}{2}\exp(-i\frac{\pi}{4})$. As a result,
$\mathcal{N}=2$; $A_1=G_2=\frac{\sqrt{2}} {2}\exp(i\frac{\pi}{4})$ and
$A_2=G_4=\frac{\sqrt{2}}{2}\exp(-i\frac{\pi}{4})$, respectively.

The pseudoclassical map Eq. \eqref{eq:keyresult} can be written as
\begin{align}
\mathcal{M}: (\Theta, \Phi) \rightarrow
\begin{cases}
(\tilde \Theta_\delta, \tilde \Phi_\delta+\pi);~~\frac{\sqrt{2}}{2}\exp(i\frac{\pi}{4}),\\
(\tilde \Theta_\delta, \tilde \Phi_\delta);~~\frac{\sqrt{2}}{2}\exp(-i\frac{\pi}{4}).
\end{cases}
\tag{C1}
\end{align}

For the particular case that $\alpha=\frac{\pi}{2}$, by taking into account
the properties of~\eqref{eq:alpha-half-pi}, we can write down the pseudoclassical
map step by step as follows. Note that the notation is the same as in
Appendix~\ref{App:B} and the constant factor $\frac{\sqrt{2}}{2}$ of $A_1$ and
$A_2$ is dropped for the sake of clearness and convenience.

\boldmath $n=0\rightarrow n=1$ \unboldmath:
\begin{align}
(\Theta(0), \Phi(0)) \rightarrow
\begin{cases}
(\tilde \Theta_\delta(1), \tilde \Phi_\delta(1));~~\exp(-i\frac{\pi}{4}),\\
(\tilde \Theta_\delta(1), \tilde \Phi_\delta(1)+\pi);~~\exp(i\frac{\pi}{4});
\end{cases}
\nonumber
\end{align}

\boldmath $n=1\rightarrow n=2$ \unboldmath:
\begin{align}
(\tilde \Theta_\delta(1), \tilde \Phi_\delta(1));~~\exp(-i\frac{\pi}{4}) \rightarrow
\begin{cases}
(\tilde \Theta_\delta(2), \tilde \Phi_\delta(2));~~-i,\\
(\tilde \Theta_\delta(2), \tilde \Phi_\delta(2)+\pi);~~1;
\end{cases}
\nonumber
\end{align}

\begin{align}
(\tilde \Theta_\delta(1), \tilde \Phi_\delta(1)+\pi);~~\exp(i\frac{\pi}{4}) \rightarrow
\begin{cases}
(\pi-\tilde \Theta_\delta(2), \pi-\tilde \Phi_\delta(2));~~1,\\
(\pi-\tilde \Theta_\delta(2), 2\pi-\tilde \Phi_\delta(2));~~i;
\end{cases}
\nonumber
\end{align}
	
\boldmath $n=2\rightarrow n=3$ \unboldmath:
\begin{align}
(\tilde \Theta_\delta(2), \tilde \Phi_\delta(2));~~-i \rightarrow
\begin{cases}
(\tilde \Theta_\delta(3), \tilde \Phi_\delta(3));~~\exp(-i\frac{3\pi}{4}),\\
(\tilde \Theta_\delta(3), \tilde \Phi_\delta(3)+\pi);~~\exp(-i\frac{\pi}{4});
\end{cases}
\nonumber
\end{align}

\begin{align}
(\tilde \Theta_\delta(2), \tilde \Phi_\delta(2)+\pi);~~1 \rightarrow
\begin{cases}
(\pi-\tilde \Theta_\delta(3), \pi-\tilde \Phi_\delta(3));~~\exp(-i\frac{\pi}{4}),\\
(\pi-\tilde \Theta_\delta(3), 2\pi-\tilde \Phi_\delta(3));~~\exp(i\frac{\pi}{4});
\end{cases}
\nonumber
\end{align}

\begin{align}
(\pi-\tilde \Theta_\delta(2), \pi-\tilde \Phi_\delta(2));~~1 \rightarrow
\begin{cases}
(\tilde \Theta_\delta(3), \tilde \Phi_\delta(3));~~\exp(i\frac{\pi}{4}),\\
(\tilde \Theta_\delta(3), \tilde \Phi_\delta(3)+\pi);~~\exp(-i\frac{\pi}{4});
\end{cases}
\nonumber
\end{align}

\begin{align}
(\pi-\tilde \Theta_\delta(2), 2\pi-\tilde \Phi_\delta(2));~~i \rightarrow
\begin{cases}
(\pi-\tilde \Theta_\delta(3), \pi-\tilde \Phi_\delta(3));~~\exp(i3\frac{\pi}{4}),\\
(\pi-\tilde \Theta_\delta(3), 2\pi-\tilde \Phi_\delta(3));~~\exp(i\frac{\pi}{4});
\end{cases}
\nonumber
\end{align}

Note that at $n=3$, the total amplitude of the point $(\tilde \Theta_\delta(3),
\tilde \Phi_\delta(3))$ vanishes and so does that of $(\pi-\tilde \Theta_\delta(3),
\pi-\tilde \Phi_\delta(3))$ as a consequence of coherence cancellation, while other
two points $(\tilde \Theta_\delta(3), \tilde \Phi_\delta(3)+\pi)$ and
$(\pi-\tilde \Theta_\delta(3), 2\pi-\tilde \Phi_\delta(3))$ remain.

\boldmath $n=3\rightarrow n=4$ \unboldmath:
\begin{align}
(\Theta(3), \Phi(3)+\pi);\exp(-i\frac{\pi}{4}) \rightarrow
\begin{cases}
(\pi-\tilde \Theta_\delta(4), 2\pi-\tilde \Phi_\delta(4));~~1,\\
(\pi-\tilde \Theta_\delta(4), \pi-\tilde \Phi_\delta(4));~~-i;
\end{cases}
\nonumber
\end{align}

\begin{align}
(\pi-\tilde \Theta(3), 2\pi-\tilde \Phi(3));\exp(i\frac{\pi}{4}) \rightarrow
\begin{cases}
(\pi-\tilde \Theta_\delta(4), \pi-\tilde \Phi_\delta(4));~~i,\\
(\pi-\tilde \Theta_\delta(4), 2\pi-\tilde \Phi_\delta(4));~~1;
\end{cases}
\nonumber
\end{align}

Note that at $n=4$, the total amplitude of $(\pi-\tilde \Theta_\delta(4),
\pi-\tilde \Phi_\delta(4))$ turns out to be zero, but the other point
$(\pi-\tilde \Theta_\delta(4), 2\pi-\tilde \Phi_\delta(4))$ survives.

\boldmath $n=4\rightarrow n=5$ \unboldmath:
\begin{align}
(\pi-\tilde \Theta_\delta(4), 2\pi-\tilde \Phi_\delta(4));1 \rightarrow
\begin{cases}
(\pi-\tilde \Theta_\delta(5), \pi-\tilde \Phi_\delta(5));~~\exp(i\frac{\pi}{4}),\\
(\pi-\tilde \Theta_\delta(5), 2\pi-\tilde \Phi_\delta(5));~~\exp(-i\frac{\pi}{4});
\end{cases}
\nonumber
\end{align}

\boldmath $n=5\rightarrow n=6$ \unboldmath:
\begin{align}
(\pi-\tilde \Theta_\delta(5), \pi-\tilde \Phi_\delta(5));~~\exp(i\frac{\pi}{4}) \rightarrow
\begin{cases}
(\tilde \Theta_\delta(6), \tilde \Phi_\delta(6));~~i,\\
(\tilde \Theta_\delta(6), \tilde \Phi_\delta(6)+\pi);~~1;
\end{cases}
\nonumber
\end{align}

\begin{align}
(\tilde \pi-\Theta_\delta(5), 2\pi-\tilde \Phi_\delta(5));~~\exp(-i\frac{\pi}{4}) \rightarrow
\begin{cases}
(\pi-\tilde \Theta_\delta(6), \pi-\tilde \Phi_\delta(6));~~1\\
(\pi-\tilde \Theta_\delta(6), 2\pi-\tilde \Phi_\delta(6));~~-i;
\end{cases}
\nonumber
\end{align}
	
\boldmath $n=6\rightarrow n=7$ \unboldmath:
\begin{align}
(\tilde \Theta_\delta(6), \tilde \Phi_\delta(6));~~i \rightarrow
\begin{cases}
(\tilde \Theta_\delta(7), \tilde \Phi_\delta(7));~~\exp(i\frac{\pi}{4}),\\
(\tilde \Theta_\delta(7), \tilde \Phi_\delta(7)+\pi);~~\exp(i3\frac{\pi}{4});
\end{cases}
\nonumber
\end{align}

\begin{align}
(\tilde \Theta_\delta(6), \tilde \Phi_\delta(6)+\pi);~~1 \rightarrow
\begin{cases}
(\pi-\tilde \Theta_\delta(7), \pi-\tilde \Phi_\delta(7));~~\exp(-i\frac{\pi}{4}),\\
(\pi-\tilde \Theta_\delta(7), 2\pi-\tilde \Phi_\delta(7));~~\exp(i\frac{\pi}{4});
\end{cases}
\nonumber
\end{align}

\begin{align}
(\pi-\tilde \Theta_\delta(6), \pi-\tilde \Phi_\delta(6));~~1 \rightarrow
\begin{cases}
(\tilde \Theta_\delta(7), \tilde \Phi_\delta(7));~~\exp(i\frac{\pi}{4}),\\
(\tilde \Theta_\delta(7), \tilde \Phi_\delta(7)+\pi);~~\exp(-i\frac{\pi}{4});
\end{cases}
\nonumber
\end{align}

\begin{align}
(\pi-\tilde \Theta_\delta(6), 2\pi-\tilde \Phi_\delta(6));~~-i \rightarrow
\begin{cases}
(\pi-\tilde \Theta_\delta(7), \pi-\tilde \Phi_\delta(7));~~\exp(-i\frac{\pi}{4}),\\
(\pi-\tilde \Theta_\delta(7), 2\pi-\tilde \Phi_\delta(7));~~\exp(-i\frac{3\pi}{4});
\end{cases}
\nonumber
\end{align}

Similar to that at $n=3$, at $n=7$, the points eventually emerge are $(\tilde \Theta_\delta(7),
\tilde \Phi_\delta(7))$ and $(\pi-\tilde \Theta_\delta(7), \pi-\tilde \Phi_\delta(7))$.
The other two vanish.
	
\boldmath $n=7\rightarrow n=8$ \unboldmath:
\begin{align}
(\Theta(7), \Phi(7));\exp(i\frac{\pi}{4}) \rightarrow
\begin{cases}
(\tilde \Theta_\delta(8), \tilde \Phi_\delta(8));~~1,\\
(\tilde \Theta_\delta(8), \tilde \Phi_\delta(8)+\pi);~~i;
\end{cases}
\nonumber
\end{align}
\begin{align}
(\pi-\tilde \Theta(7), \pi-\tilde \Phi(7));\exp(-i\frac{\pi}{4}) \rightarrow
\begin{cases}
(\tilde \Theta_\delta(8), \tilde \Phi_\delta(8)+\pi);~~1,\\
(\tilde \Theta_\delta(8), \tilde \Phi_\delta(8)+\pi);~~-i.
\end{cases}
\nonumber
\end{align}

Finally, after eight kicks, the pseudoclassical dynamics brings the initial condition
to $(\tilde \Theta_\delta(8), \tilde \Phi_\delta(8))$, the same as the classical
map does. We thus have Eq.~\eqref{eq:caseIIJ} immediately.

\end{document}